\documentclass{article}

\usepackage{arxiv}

\usepackage[utf8]{inputenc} 
\usepackage[T1]{fontenc}    
\usepackage{hyperref}       
\usepackage{url}            
\usepackage{booktabs}       
\usepackage{amsfonts}       
\usepackage{nicefrac}       
\usepackage{microtype}      
\usepackage{lipsum}		
\usepackage{graphicx}
\usepackage[square,sort,comma,numbers]{natbib}
\usepackage{doi}
\usepackage{subcaption}

\title{Analysis of Energy Consumption in a Precision Beekeeping System}

\date{October 2020}

\author{ \href{http://perso.ens-lyon.fr/hugo.hadjur/}{Hugo Hadjur}\thanks{\url{http://perso.ens-lyon.fr/hugo.hadjur/}} \\
	em\textbf{lyon business school}\\
	Inria, Univ Lyon, EnsL, UCBL, CNRS, LIP\\
	Lyon, France \\
	\texttt{hugo.hadjur@ens-lyon.fr} \\
	\And
	\href{https://scholar.google.com/citations?user=Wh-CURQAAAAJ}{Doreid Ammar}\thanks{\url{https://scholar.google.com/citations?user=Wh-CURQAAAAJ}} \\
	em\textbf{lyon business school}\\
	Ecully, France \\
	\texttt{ammar@em-lyon.com} \\
	\And
	\href{https://perso.ens-lyon.fr/laurent.lefevre/}{Laurent Lefevre}\thanks{\url{http://perso.ens-lyon.fr/laurent.lefevre/}} \\
	Inria, Univ Lyon, EnsL, UCBL, CNRS, LIP\\
	Lyon, France \\
	\texttt{laurent.lefevre@ens-lyon.fr} \\
}

\renewcommand{\headeright}{}
\renewcommand{\undertitle}{}
\renewcommand{\shorttitle}{}

\hypersetup{
pdftitle={Analysis of Energy Consumption in a Precision Beekeeping System},
pdfsubject={q-bio, cs.AR},
pdfauthor={Hugo Hadjur, Doreid Ammar, Laurent Lefevre},
pdfkeywords={internet of things, energy consumption, benchmarking, smart agriculture, precision beekeeping},
}

\begin{document}
\maketitle

\begin{abstract}
Honey bees have been domesticated by humans for several thousand years and mainly provide honey and pollination, which is fundamental for plant reproduction.
Nowadays, the work of beekeepers is constrained by external factors that stress their production (parasites and pesticides among others).
Taking care of large numbers of beehives is time-consuming, so integrating sensors to track their status can drastically simplify the work of beekeepers.
Precision beekeeping complements beekeepers' work thanks to the Internet of Things (IoT) technology. If used correctly, data can help to make the right diagnosis for honey bees colony, increase honey production and decrease bee mortality.
Providing enough energy for on-hive and in-hive sensors is a challenge. Some solutions rely on energy harvesting, others target usage of large batteries. Either way, it is mandatory to analyze the energy usage of embedded equipment in order  to design an energy efficient and autonomous bee monitoring system.
This paper relies on a fully autonomous IoT framework that collects environmental and image data of a beehive.
It consists of a data collecting node (environmental data sensors, camera, Raspberry Pi and Arduino) and a solar energy supplying node. Supported services are analyzed task by task from an energy profiling and efficiency standpoint, in order to identify the highly pressured areas of the framework. This first step will guide our goal of designing a sustainable precision beekeeping system, both technically and energy-wise.
\end{abstract}


\section{Introduction}
Whether it is used by companies or individuals, technology applied to agriculture (smart agriculture or precision agriculture) has brought a whole new dimension to farming. With the new modern constraints of growing demand and farming area shrinking, the need is no longer to expand, but to optimize, while taking into consideration the environmental challenges. To do so, farmers nowadays make use of IoT technologies that collect relevant data regarding their plantations and data-related tools to optimize their work.

Recently, several stress factors emerged from the agricultural industry and apiculture, among others, is affected. Precision beekeeping is a branch of precision agriculture linked to apiculture.
The goal is to support beekeepers so that crucial events are detected as early as possible. Such events may include the swarming state, queenless state, infected colony, attacked colony and dead colony.
Designing a precision beekeeping system deployed into wild zones usually requires efficient use of the energy budget, which is mainly provided by batteries.


Monitoring and profiling energy consumption of large scale systems is a real challenge~\cite{OAL14}.
The objective of this work is to observe and analyze the energy consumption of all deployed hardware, software and services components of a precision beekeeping system.
Through this analysis, we can embed all costs of specific steps (idle mode, boot up, shutdown, stress) combined with required services (sensors data collection, image capture, data exchanges) and propose a complete energy consumption overview.


The paper is organized as follows: 
Section~\ref{soa} provides a brief overview of selected previous works addressing precision beekeeping.
In Section~\ref{beekeeping}, we give an overview of the precision beekeeping system, describe the data collection process, and present the experimental setup used to analyze the energy consumption. Section~\ref{bench} focuses on the analysis of energy consumption of some specific steps (idle, boot, shutdown) while in Section~\ref{energy} we present an analysis of the energy consumption of the beekeeping services.  In Section~\ref{temperature}, we analyze the energy consumption of the system in several temperature conditions. Section~\ref{conclusion} concludes the paper and shares directions for future work.

\section{Background and Related work}
\label{soa}
In recent years, several studies have underlined the potential of integrating intelligent digital technologies for monitoring honey-bees.
Edwards-Murphy et al. propose in~\cite{bWSN16} a bee monitoring system
to describe the internal conditions of a beehive using data from sensors that are embedded within the beehive. The collected data are: temperature, humidity, Carbon Dioxide (CO$_2$) levels, and Oxygen (O$_2$). A machine model is developed in~\cite{bWSN16} to detect a subset of beehive status
and alert the beekeeper when an important beehive change is detected.
An audio-based monitoring system for detecting the presence of the queen bee inside the beehive 
is introduced in~\cite{audio-beehiveStates18}, using data from the NU-Hive project~\cite{cecchi2018a}. The proposed system exploits the sound emitted by the beehives
to determine the presence of the queen bee inside the beehive.
Similarly, audio-based monitoring systems are also used to detect a beehive state known as "swarming", such as the work conducted in~\cite{FERRARI200872}. 
Although beekeepers who decide to integrate sensors to their beehives still heavily rely on weight (to be correlated with honey quantity), companies release monitoring systems capable of tracking real time temperature of the inside of beehives, humidity, sound and gas levels.
A variety of products is available on the market. Label Abeille\footnote{\url{https://www.label-abeille.org}} and  Arnia\footnote{\url{https://www.arnia.co.uk}} cover a wide range of metrics outside the beehive (temperature, humidity, weight of the beehive and other environmental data). Mellisphera\footnote{\url{https://www.mellisphera.com}} allows monitoring temperature and humidity from inside the beehive. Pollenity\footnote{\url{https://pollenity.com}} develops a sensor that can be attached to an in-hive frame and collect temperature, humidity and sound up close from the bees.
\\The paper from Ammar et al.~\cite{Ammar} introduces the system which is analyzed in this paper.
Here, an in-depth analysis of the energy consumption of each step of the precision beekeeping system is conducted, in order to design an autonomous, energy-efficient and sustainable connected beehive.

\section{Precision Beekeeping System}
\label{beekeeping}
\begin{figure}[htbp]
\centering
\includegraphics[width=0.6\columnwidth]{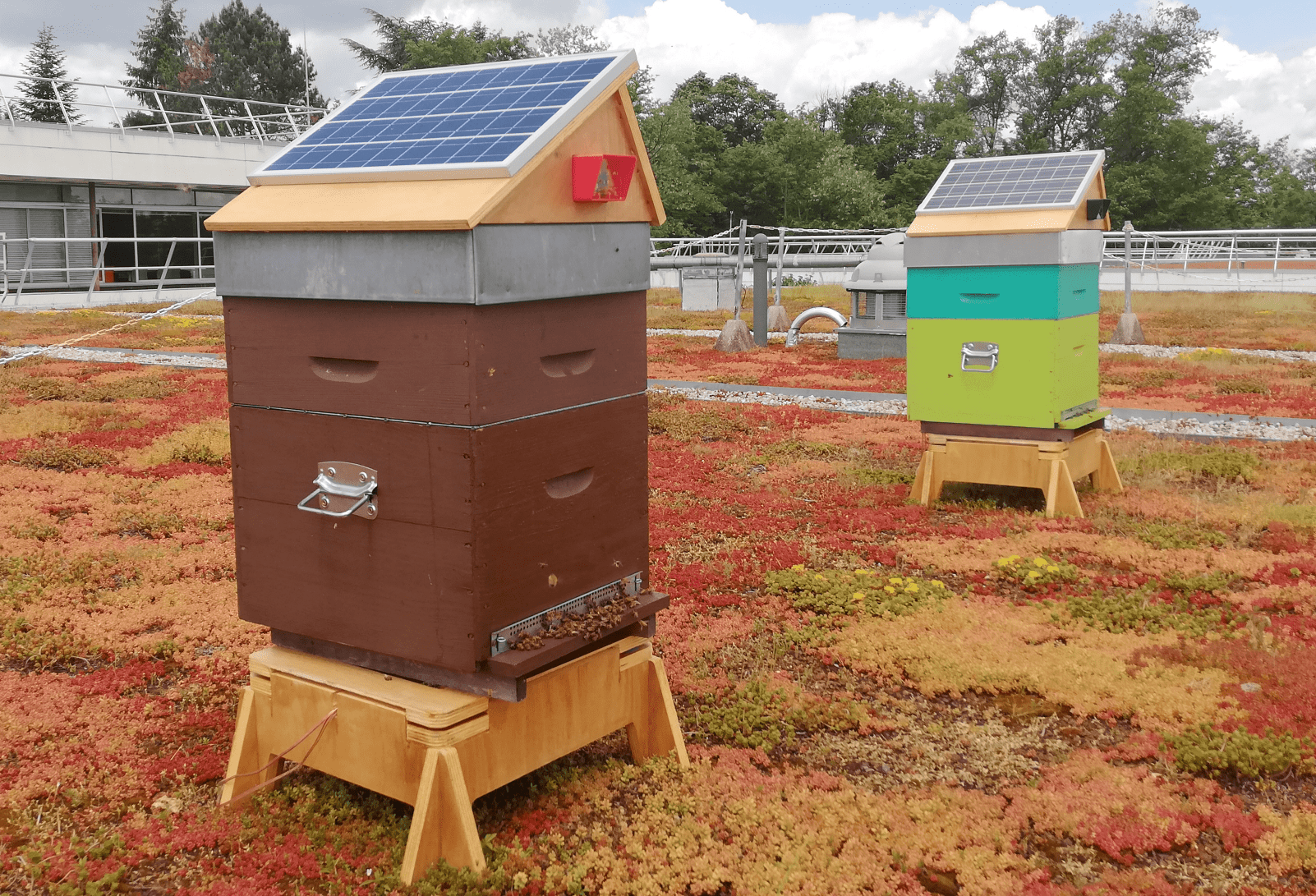}
\caption{Two precision beekeeping systems deployed on site}\label{fig:overview}
\end{figure}

This paper aims to analyze the energy consumption of a precision beekeeping system built at em\textbf{lyon business school} and previously described in~\cite{Ammar}. The name of the system (beehive equipped with IoT components) is the ``Makers' Beehives''.
Figure~\ref{fig:overview} shows a picture of two Makers' Beehives that are deployed on the roof of em\textbf{lyon business school}. 

In this section, we give a brief overview of the architecture and the electronic components of a Makers' Beehive, as well as, the data acquisition process.
Following that, we elaborate on the experimental setup used to analyze the energy consumption of this precision beekeeping system.

\subsection{System Overview of a Makers' Beehive}
A Makers' Beehive consists of a commonly used Dadant beehive to which two extra parts are added, namely, a wooden base and a solar roof that are placed respectively under and on top of the beehive.
The base functions as both a stand for the beehive and a scale thanks to four load sensors located at the corners of the base frame.
The roof hosts the electronics, and is equipped with a solar panel tilted to an optimum angle for maximum annual energy production.
More specifically, the electronic components are powered by a lithium-ion polymer (LiPo) battery with the capacity of 33000 mAh and a maximum voltage of 5 V. The battery is connected via a DC/DC step-down converter to a 25 W polycrystalline solar panel.
The IoT nodes of the Makers' Beehive (Figure \ref{fig:setup}) consist of: a 5 V Adafruit Pro Trinket, a Raspberry Pi 3 Model B micro-computer, a Raspberry Pi Camera, and an Adafruit METRO 328 based on the Atmel ATmega328 single-chip micro-controller~\cite{Ammar}.
\subsection{Data Acquisition in a Makers'  Beehive}\label{steps}
The data acquisition process in a Makers' Beehive consists of the following steps :
\begin{description}
\item[\textit{Step 0:}] The Trinket turns ON and the boot up of the Makers' Beehive system is done.
\item[\textit{Step 1:}] The Raspberry Pi runs a Python script that collects the measurements (i.e., temperature, humidity, noise, light, Carbon monoxide (CO), Nitrogen dioxide (NO$_2$), and the beehive's weight) from various sensors on the METRO 328.
\item[\textit{Step 2:}] The Raspberry Pi camera captures a series of 20 top view photos of the beehive entrance (one per second).
\item[\textit{Step 3:}] The captured pictures are converted to a single gif file that is uploaded to an online image sharing website.
\item[\textit{Step 4:}] The collected measurements and the url of the uploaded gif file are uploaded to a remote server. By the end of this step, a software update check is performed and the \textit{git pull} command is executed when an outdated software is detected.
\item[\textit{Step 5:}] The system is granted a time interval of 5 minutes to accomplish the data acquisition process. Once the data is successfully uploaded, the Raspberry Pi shuts off completely. 
However, if the data acquisition process is not accomplished within the granted time, the Trinket turns OFF leading to a system shutdown.
\end{description}

The Makers' Beehive periodically executes sensing cycles.
Every hour, the system wakes up, performs the data acquisition process and then goes into sleep mode again.
It is also worth noting that the data acquisition steps are performed over a short time interval of 1 to 5 minutes.

The following sections describe an experimental replica of the Makers' Beehive and its power consumption in various situations.

\subsection{Testbed for Monitoring and Profiling the Energy Consumption of IoT nodes}
To measure the energy consumption of our precision beekeeping system we developed an indoor experimental setup capable of monitoring and profiling the energy consumption of the IoT nodes.
\begin{figure}[htbp]
\centering
\includegraphics[width=0.6\columnwidth]{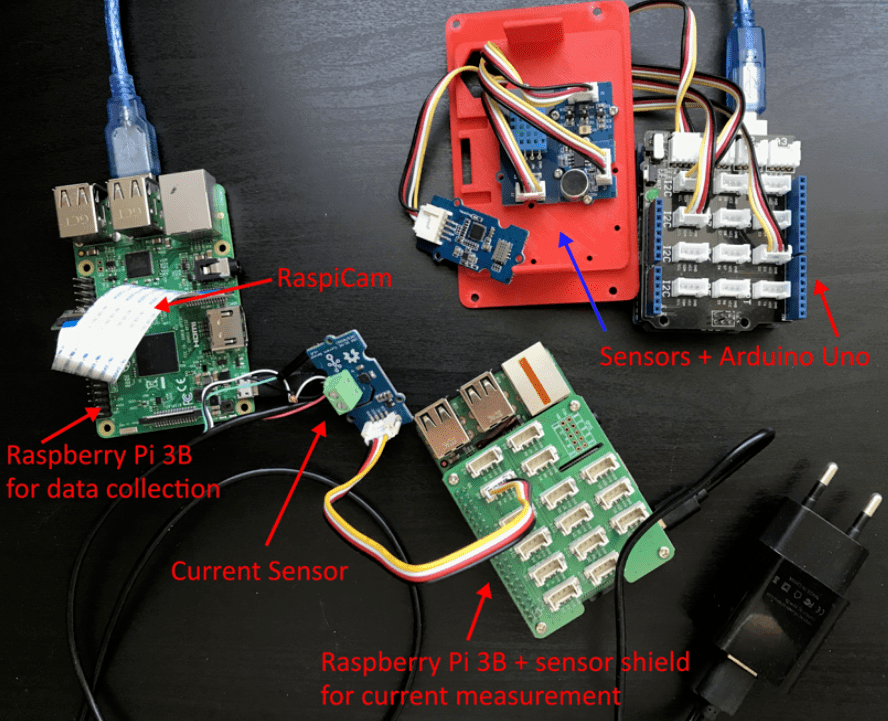}
\caption{Experimental setup for monitoring energy consumption\label{fig:setup}}
\end{figure}

Figure~\ref{fig:setup} shows an overview of our testbed.
The testbed consists of a duplicated version of the sensor nodes that are responsible for performing the data acquisition process in the Makers' Beehive.
However, the solar panel, converter, battery and Trinket were removed and replaced by a microUSB power connector that is directly connected via a 5V voltage regulator to a power plug.
Furthermore, we make use of an additional Raspberry Pi 3 Model, a Base Hat for Raspberry Pi and a $\pm$5A DC/AC Current Sensor (ACS70331).
The current sensor is used to measure the power consumption of the Makers' Beehive. This sensor is connected to the new Raspberry Pi via the Base Hat.


\section{Benchmarking the energy consumption of the precision beekeeping system}
\label{bench}
We first observe the energy consumption of specific conditions and steps (idle, boot, shutdown and stress) in order to analyze several costs.

\begin{figure}[h]
\centering
\includegraphics[width=0.6\columnwidth]{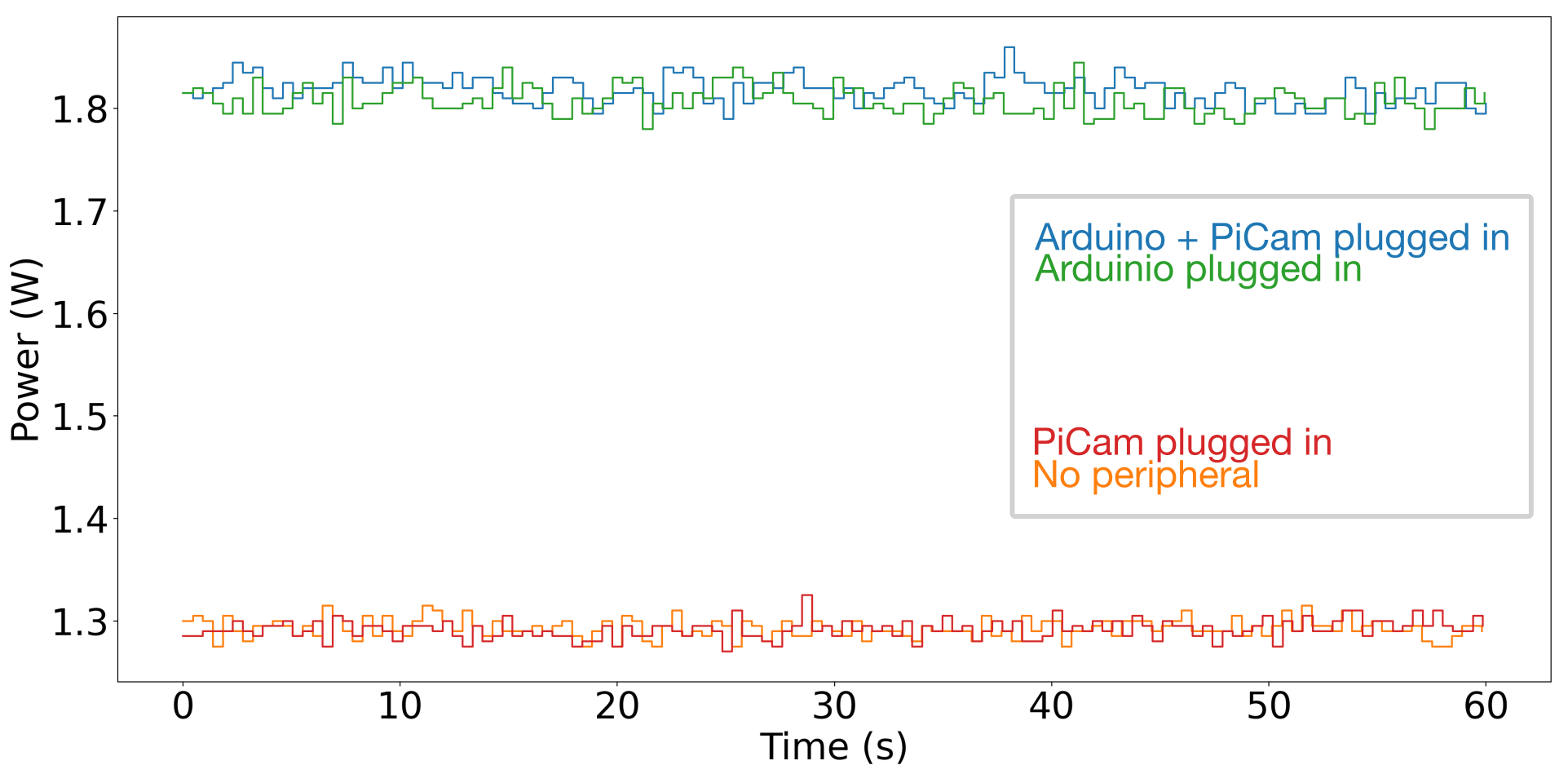}
\caption{Power consumption of a Raspberry Pi in Idle mode over a 60-second time window}
 \label{fig:idle}
\end{figure}

\subsection{Idle Mode}

A first step toward analyzing the energy consumption of the setup and its tasks is to understand the fixed costs of the hardware. The consumption of the data acquiring Raspberry Pi was recorded while in Idle mode (Raspberry Pi logged in, the operating system launched and at a resting state). 
This experiment repeats itself four times, to determine the power consumption of the peripherals. First, the full setup is tested (Arduino sensor node plugged in and Pi Camera plugged in). Then, each peripheral is plugged in alone. Finally, the Idle consumption of the setup without any peripheral is recorded.

Figure~\ref{fig:idle} shows the extra 0.5W of power cost by the connection of the Arduino sensor node via USB. It is important to note that the energy impact of the camera peripheral (not in use) seems negligible.

\subsection{Boot and Shutdown Profiling}
\begin{figure}[htbp]
\centering
\includegraphics[width=0.6\columnwidth]{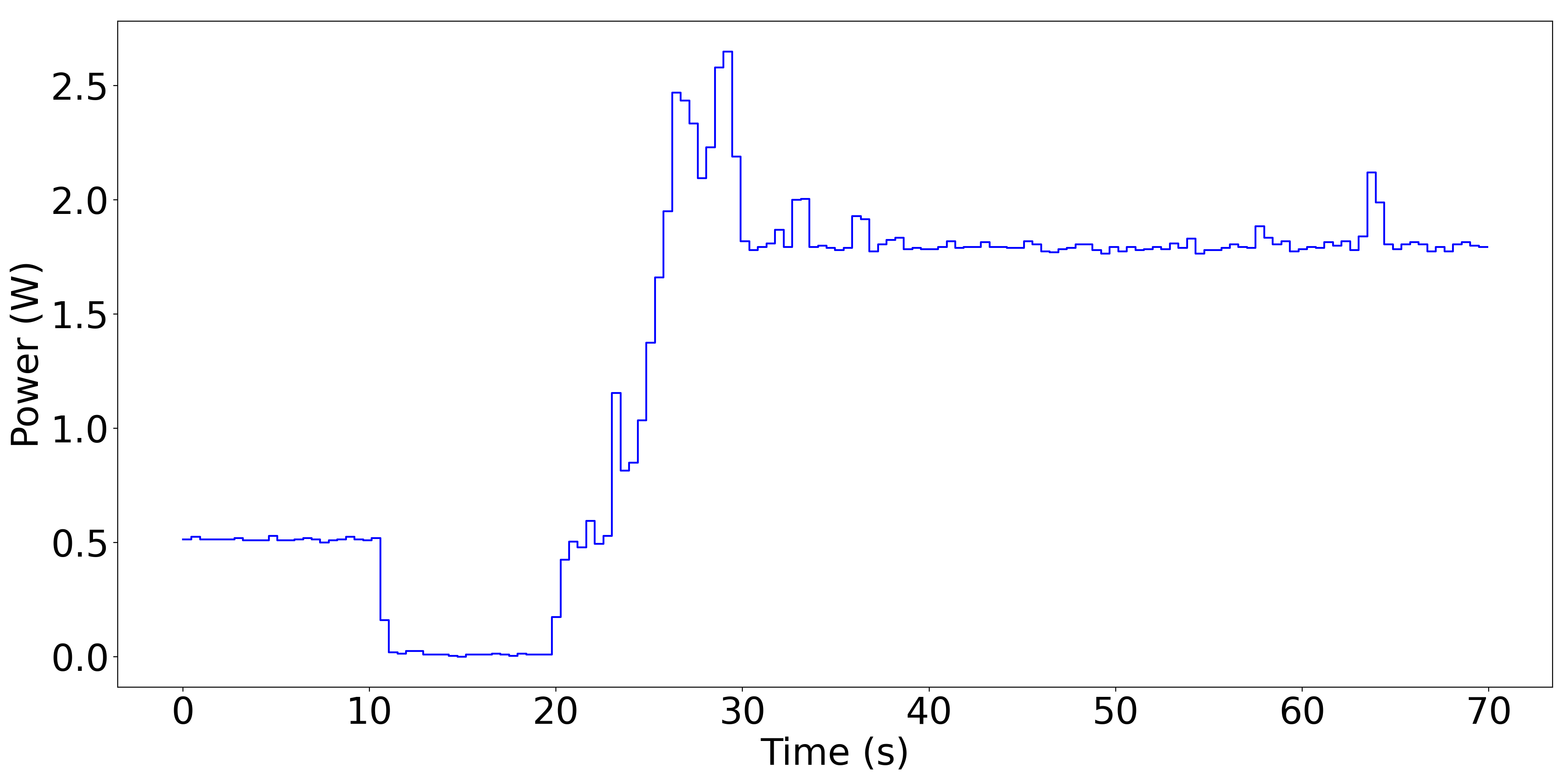} 
\caption{Power consumption of the full setup (Arduino plugged in) during boot up}
\label{fig:boot_on}
\end{figure}

\begin{figure}[htbp]
\centering
\includegraphics[width=0.6\columnwidth]{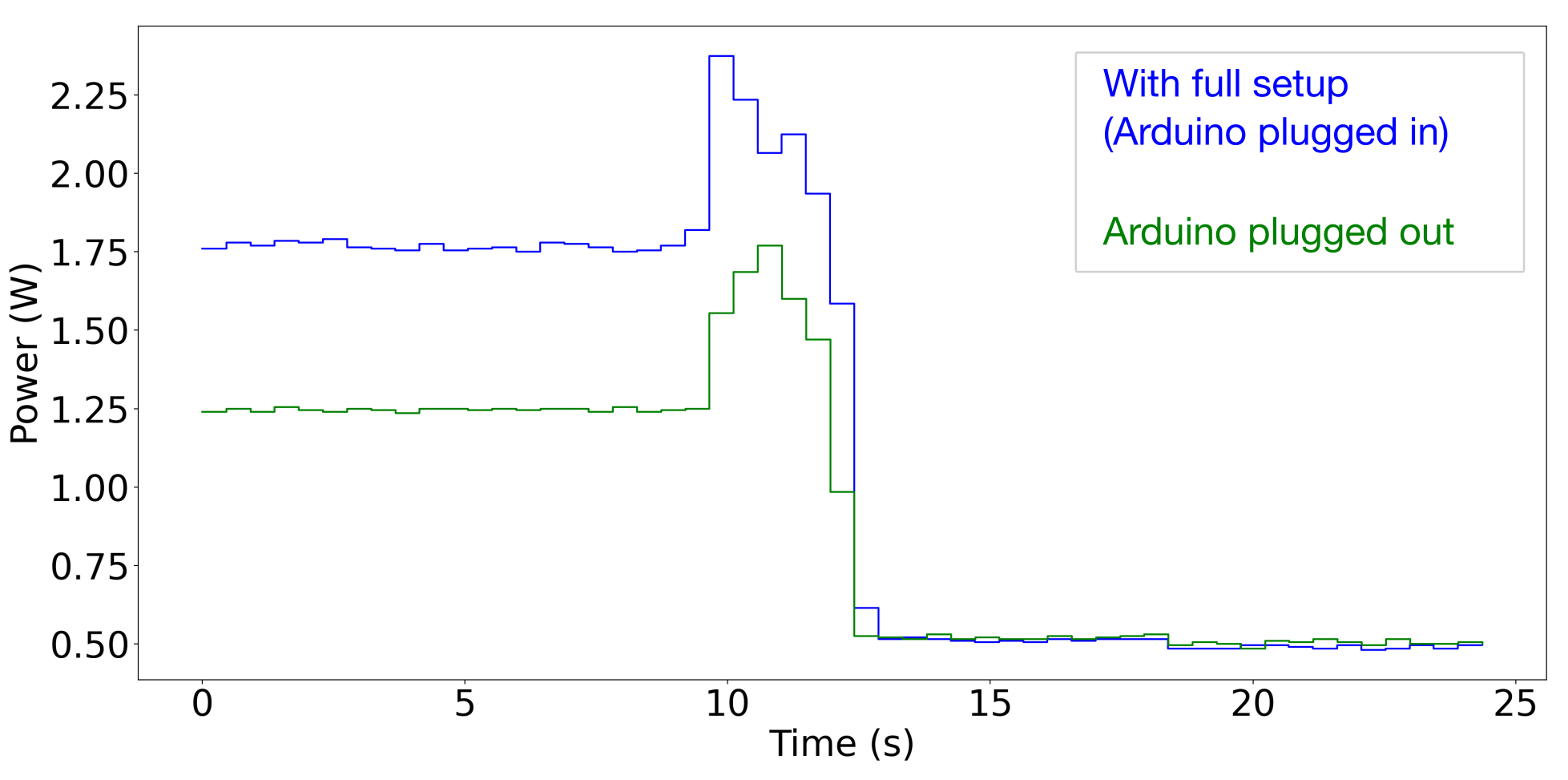}
\caption{Power consumption of the full setup -- Arduino plugged in -- (blue) and Arduino plugged out (green) during shutdown}
\label{fig:boot_shutdown}
\end{figure}

Since the in-hive system wakes up every hour, it is mandatory to analyze the details of each boot up and shutdown phase. Figures \ref{fig:boot_on} and \ref{fig:boot_shutdown} show the power evolution of a standard boot up phase (with Arduino plugged in) and the comparison of two shutdown phases.

\paragraph{Observing Boot up step} 

For the boot up phase, the recorded Raspberry Pi was turned off (0s-10s window in Figure \ref{fig:boot_on}), plugged out of the power supply (10s-20s) and plugged back in (20s mark).

The boot up phase, thanks to its quickness (10 seconds, between the 20s and 30s mark on Figure \ref{fig:boot_on}), does not require much energy (14.5 Joules in the case of Figure~\ref{fig:boot_on}): the time window where the power is greater than the resting phase is only 4-second long (26-30s).

\paragraph{Observing Shutdown step} 

For the shutdown phase, a graceful shutdown approach was used. During a graceful shutdown phase, all running processes are sent a shutdown message. It is only once those processes are closed that the interface is turned off and the filesystems unmounted (described in~\cite{Amirtharaj2019}).
A forced shutdown alternative also exists. It does not notify the running processes of the incoming shutdown. In our case, both were tested and although the forced approach turns off the system in less than one second, the small gain of energy is not worth the risk of potential memory errors. The presented results focus on the graceful approach.

The results of Figure~\ref{fig:boot_shutdown} follow the previous section findings, as the extra USB peripheral adds a fixed cost of energy while turned on. If the Raspberry Pi is turned off and still plugged in, there is no impact of the Arduino. The slight peak of power usage at the moment of the shutdown (10s mark) is explained by the messages sent to running processes  and the stop of these processes.

\subsection{Stress Testing}

To better understand the limits of the services of a Raspberry Pi 3, its performances in extreme situations are analyzed. Stress testing aims at pushing a CPU to its full power.

\begin{figure}[htbp]
\centering
\includegraphics[width=0.6\columnwidth]{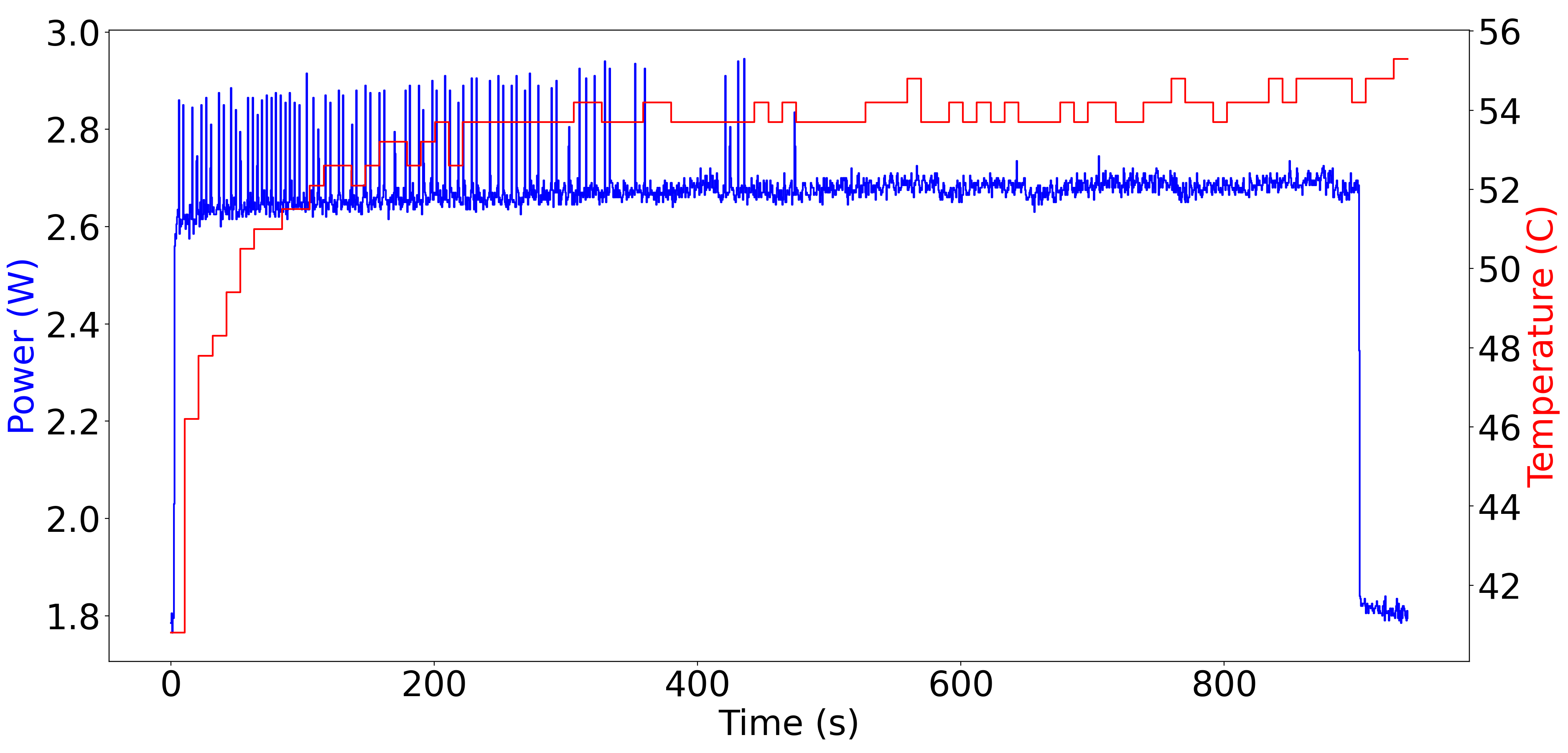}
\caption{Power consumption of stress task}
\label{fig:stress_test}
\end{figure}

\begin{figure}[htbp]
\centering
\includegraphics[width=0.6\columnwidth]{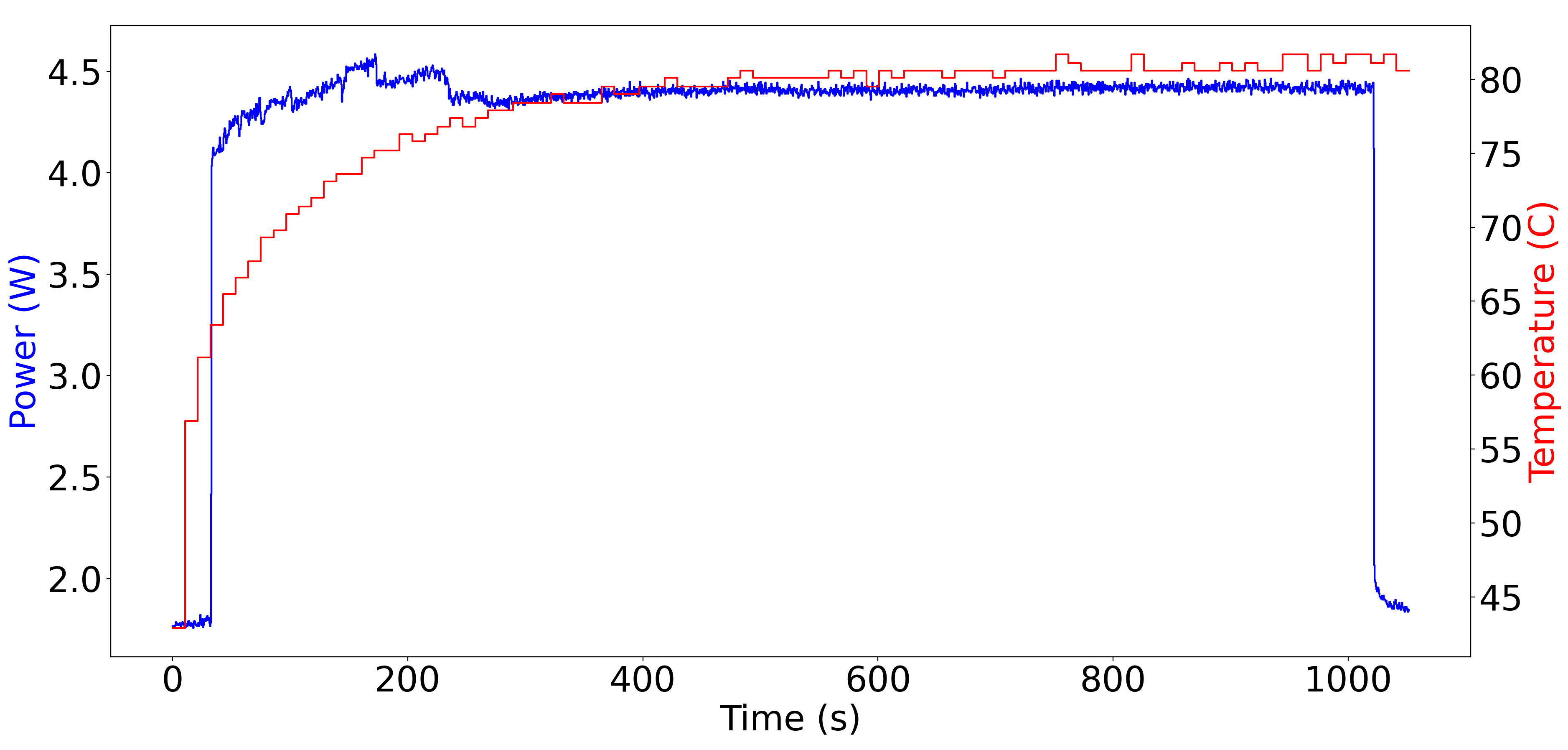}
\caption{Power consumption of CPU burn task}
\label{fig:cpu_burn}
\end{figure}

\paragraph{Experimental Conditions}

In this section, we benchmark two stress tasks, both available on a Raspbian Lite operating system:
\begin{itemize}
 \item Stress (\texttt{stress}): here, the CPU is run at its full power, all the while controlling the temperature of the chip. 
 \item CPU Burn (\texttt{cpuburn-a53}): here, the Raspberry Pi is maxed out completely without any restriction.
\end{itemize}

\paragraph{Results}

Figure~\ref{fig:stress_test} and Figure~\ref{fig:cpu_burn} show the temperature and power consumption for the two tasks along several minutes. The absolute maximal power used from the Raspberry Pi 3 lies around 4.5W, whereas the safer approach only requires between 2.6W and 2.7W, which is comparable to the energy of some tasks performed in the field.

\section{Analyzing The Energy Consumption of Beekeeping services}
\label{energy}
The second part of our analysis concerns the measurement of the energy consumption of the data acquisition process that is detailed in Section~\ref{steps}.
More precisely, we profile the power usage of the collected measurements and compute the consumed energy (\textit{Step 1}: 6.2 Joules on average), captured images (\textit{Step 2}: 23.7 Joules on average), resized and converted images (\textit{Step 3}: 126.1 Joules on average), and transmitted data (\textit{Step 4}: 13.5 Joules on average).
This analysis relies on the embedded Python code that runs inside the beehive, every hour.

\begin{figure}[htbp]
\centering
\includegraphics[width=0.6\columnwidth]{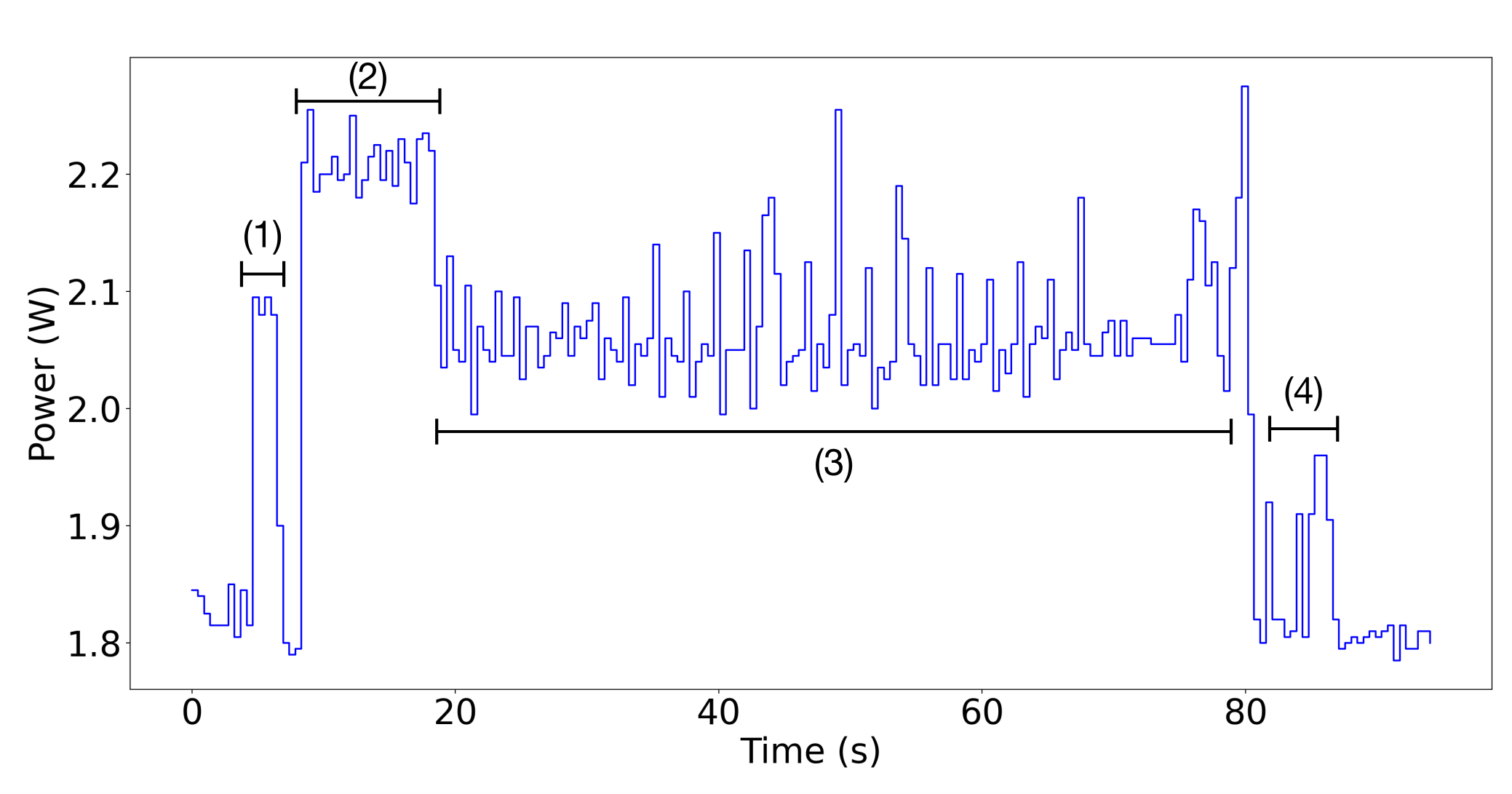}
\caption{Power usage graph of one execution of the in-hive script}
\label{fig:global_power}
\end{figure}


Figure~\ref{fig:global_power} shows the overall consumption of one execution of the in-hive script.
It is mainly divided into four energy phases, one for each step, allowing us to estimate the power cost of each.
The most energy-consuming phase is the longest: the third one (20s-80s window) which converts images into a GIF and uploads it. The GIF file, which is then uploaded to the dashboard website\footnote{\url{https://makersbeehives.herokuapp.com}} allows beekeepers to inspect the entrance of their beehive at a glance. 

An improved future version of the script should take care of the speed of the network, in order to balance the time of upload and the time of conversion, with the ultimate goal of shortening the whole process to decrease the energy consumption of the system.

\subsection{Collecting data from sensors}

The temperature, humidity, noise, light, Carbon monoxide (CO) and Nitrogen dioxide (NO2) are captured at the beginning of the hourly script. It usually takes 1 to 2 seconds to perform this step (step 1  of Figure \ref{fig:global_power}).


\subsection{Image Capturing}

The series of captured images is represented in step 2 of Figure~\ref{fig:global_power}.

There is a challenge for choosing the quantity and the resolution of images.
The resolution affects the performances of deep learning computer vision algorithms: the better the quality, the higher the efficiency of the learning model. The process of training such a model is also linked to the size of the training set: in general, a large number of images as input is able to train a deep learning model well. But at the same time, capturing and transferring a large number of high-quality images is time-consuming and CPU-consuming. 
Our goal is to find the right balance between the energy cost of the system and the quality of recommendations to beekeepers.

The Raspberry Pi camera (Raspberry Pi Camera Rev 1.3\footnote{\url{https://www.raspberrypi.org/documentation/hardware/camera/}}) and its Python module allow to pre-select the resolution of the still pictures so that there is no need to uselessly take high-resolution pictures to then downsize them. It was the case of the first version of the code which was running in the Makers' Beehive. Since then, this was changed to a better approach.

In this section, the cost of time and energy of capturing different quantities of images at different resolutions was analyzed.

\paragraph{Experimental Conditions}
Once a picture is taken, the program loops without any pause and directly runs the capture of the next one. For consistency, the camera was always placed at the same spot and with the same angle of view.

The first experiment consists of measuring the time to capture different batch sizes of 800$\times$600px images. The batches are made from 1 to 50 images, are each one at least repeated 6 times.

The second experiment focuses on the time taken to capture 10-image batches at different resolutions (list of resolutions: 640$\times$480px, 800$\times$600px, 960$\times$720px, 1024$\times$768px, 1280$\times$960px, \hspace{2mm}1400$\times$1050px, \hspace{2mm}1440$\times$1080px, \hspace{2mm}1600$\times$1200px,\\1856$\times$1392px, 1920$\times$1440px, 2048$\times$1536px and 2592$\times$1944px).
For each image resolution, the experiment of capturing a 10-image batch was repeated 12 times.

\begin{figure}[tbh]
    \centering
    \begin{subfigure}[b]{0.28\textwidth}
        \includegraphics[width=\textwidth]{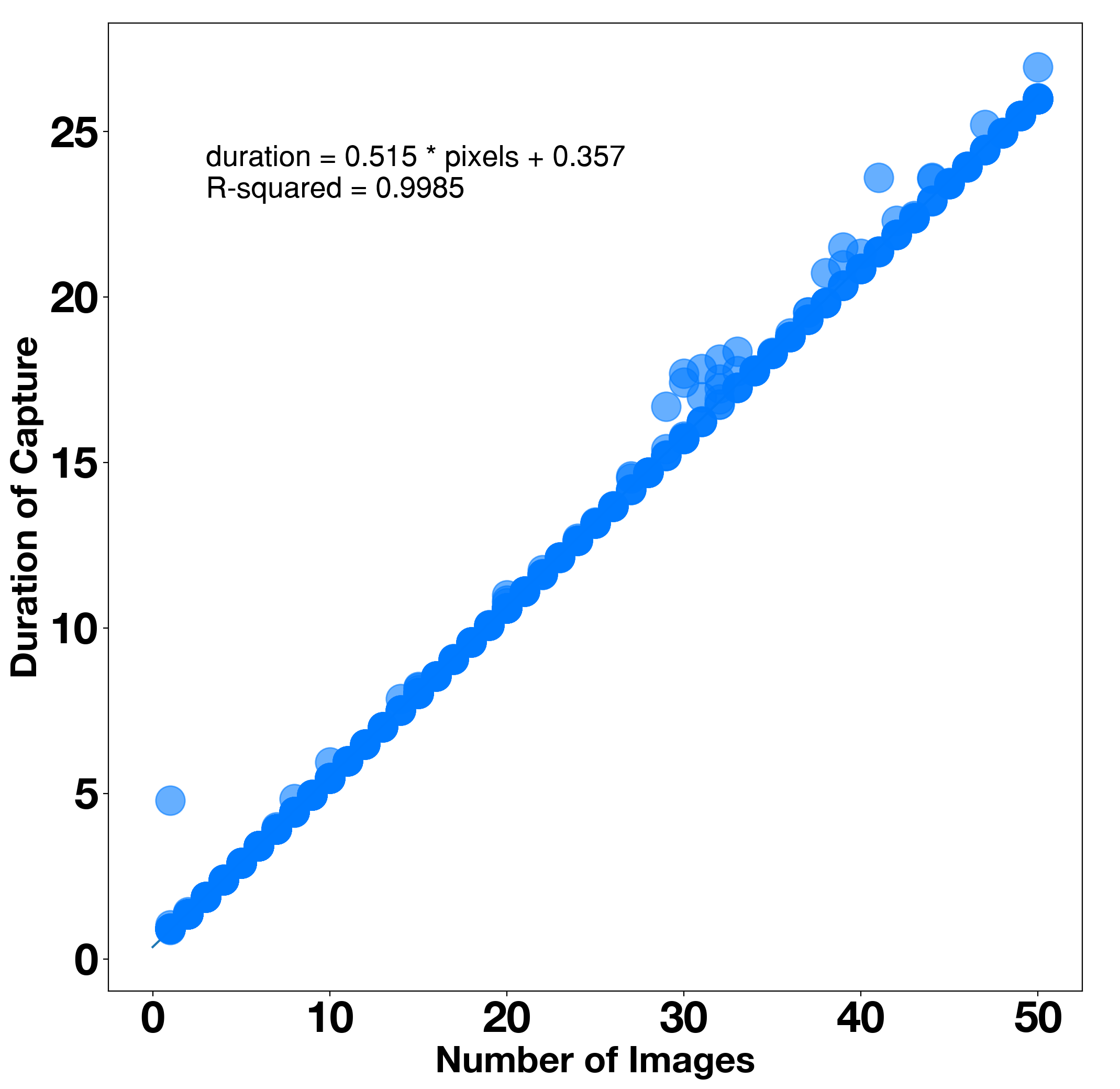}
        \caption{\footnotesize{Different batch sizes}}
        \label{fig:nb_img}
    \end{subfigure}
    ~
    \begin{subfigure}[b]{0.28\textwidth}
        \includegraphics[width=\textwidth]{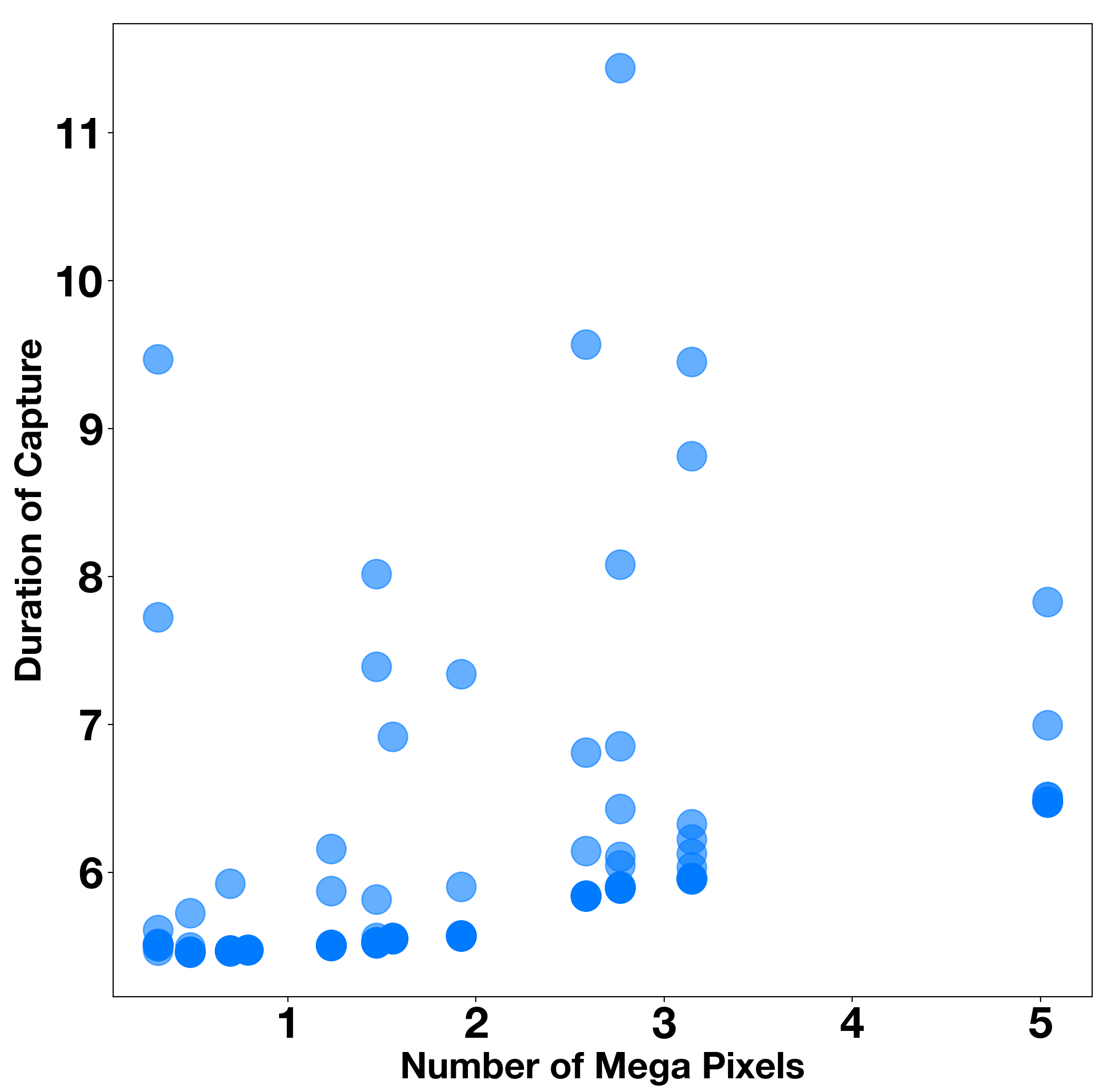}
        \caption{\footnotesize{Different resolutions}}
        \label{fig:res_img}
    \end{subfigure}
    \caption{Time needed by the PiCamera  to capture (a) different batch sizes of 800$\times$600px images and (b) 10 images with different resolutions.}\label{fig:img}
\end{figure}

\paragraph{Results}
Figure~\ref{fig:nb_img} shows that the average time per image is 0.515s, which is also the fastest frame rate at this resolution when taking pictures (the camera module also allows video capturing, not treated in our work). 

The results presented in Figure~\ref{fig:res_img} underline the disability of the camera module to consistently capture image at its optimal speed. This irregularity aside, there is also a non-linearity of the relationship between the duration of capture and the resolution of images.

%

\subsection{Network Testing}

The flow of data transmitted in the sensor network is bidirectional. On one hand, the Raspberry Pi updates the software at each iteration of the main loop: a download triggered by a \texttt{git pull} command (step 4 of Figure~\ref{fig:global_power}). On the other, data from environmental sensors and the GIF file are uploaded to a server, where it can be manually accessed (end of step 3 and step 4 of Figure~\ref{fig:global_power}). 

The goal of this section is to profile the data transmission energetically for both wired and wireless connexion types. The energy consumption of such scenarios was analyzed thanks to the iPerf3 tool~\cite{iperf}, which allows transferring network packets at the maximum achievable bandwidth.

\paragraph{Experimental Conditions}

The replica of the in-hive Raspberry Pi was connected to a local network, located in Lyon, France (Internet provider: SFR). This network has a maximum download speed of around 190Mbits/s and a maximum upload speed of around 20Mbits/s. The Ethernet cable has a maximum speed of 100Mbits/s. The server which was used to test sending and receiving data is a public server located at \url{bouygues.iperf.fr}. Its bandwidth is 10Gbits/s.

Once the Raspberry Pi connected to the network (either through Ethernet or Wi-Fi), a 50MB upload was tested thanks to the command \texttt{iperf3 -c bouygues.iperf.fr --bytes 50M}, as well as the download equivalent (\texttt{-R} added at the end of the command). 50MB is enough to get at minimum a few seconds of execution for the fastest rate of transfer (downloading through Ethernet). 
The Internet transport protocol used is TCP (Transmission Control Protocol) and there is no bandwidth restriction.

\begin{figure}[htbp]
\centering
\includegraphics[width=0.6\columnwidth]{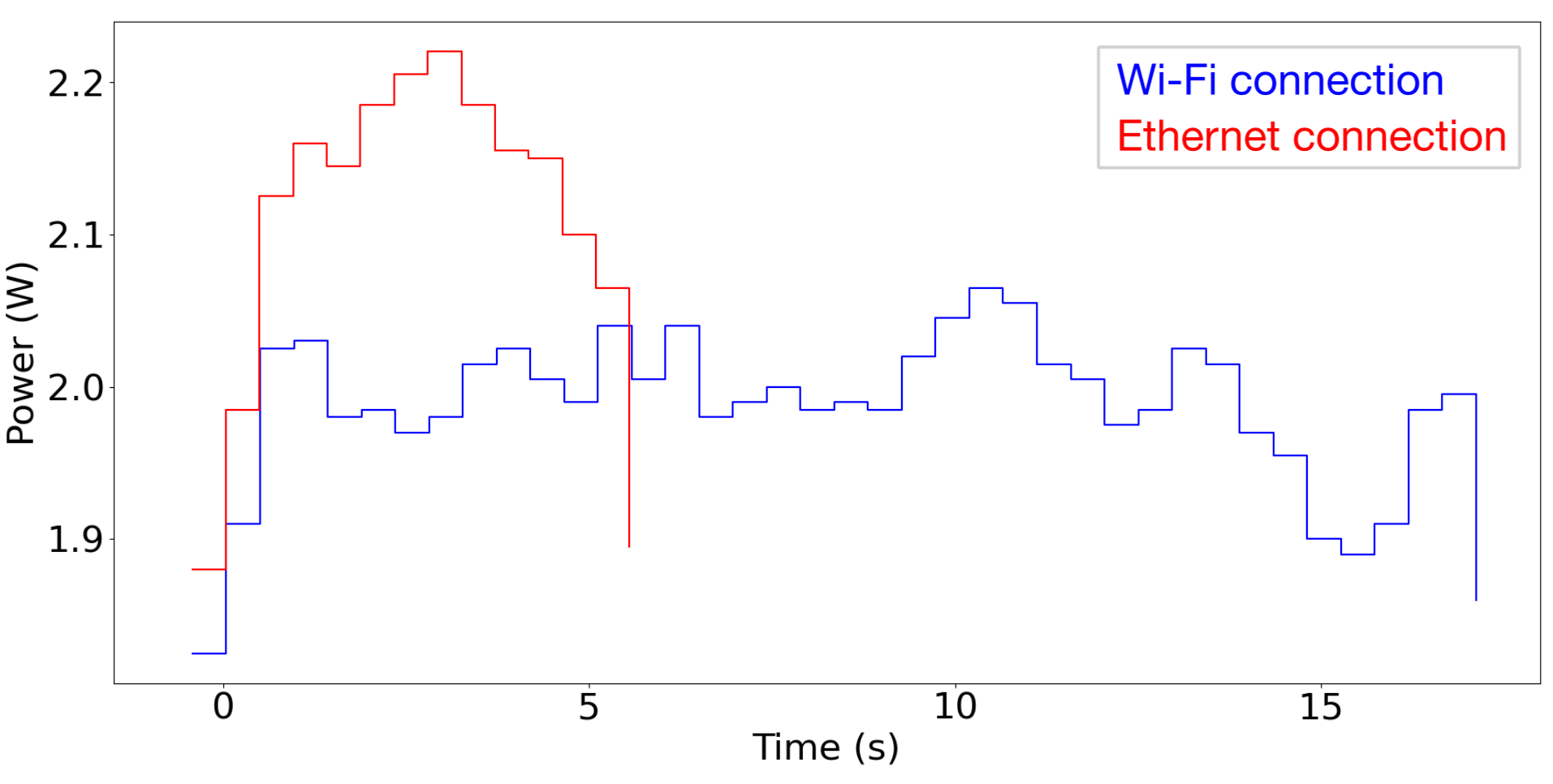}
\caption{Power consumption of the full setup downloading 50MB from a distant server. Two types of connectivity were tested: Wi-Fi (blue) and ethernet (red).} 
\label{fig:iperf_download}
\end{figure}

\begin{figure}[htbp]
\centering
\includegraphics[width=0.6\columnwidth]{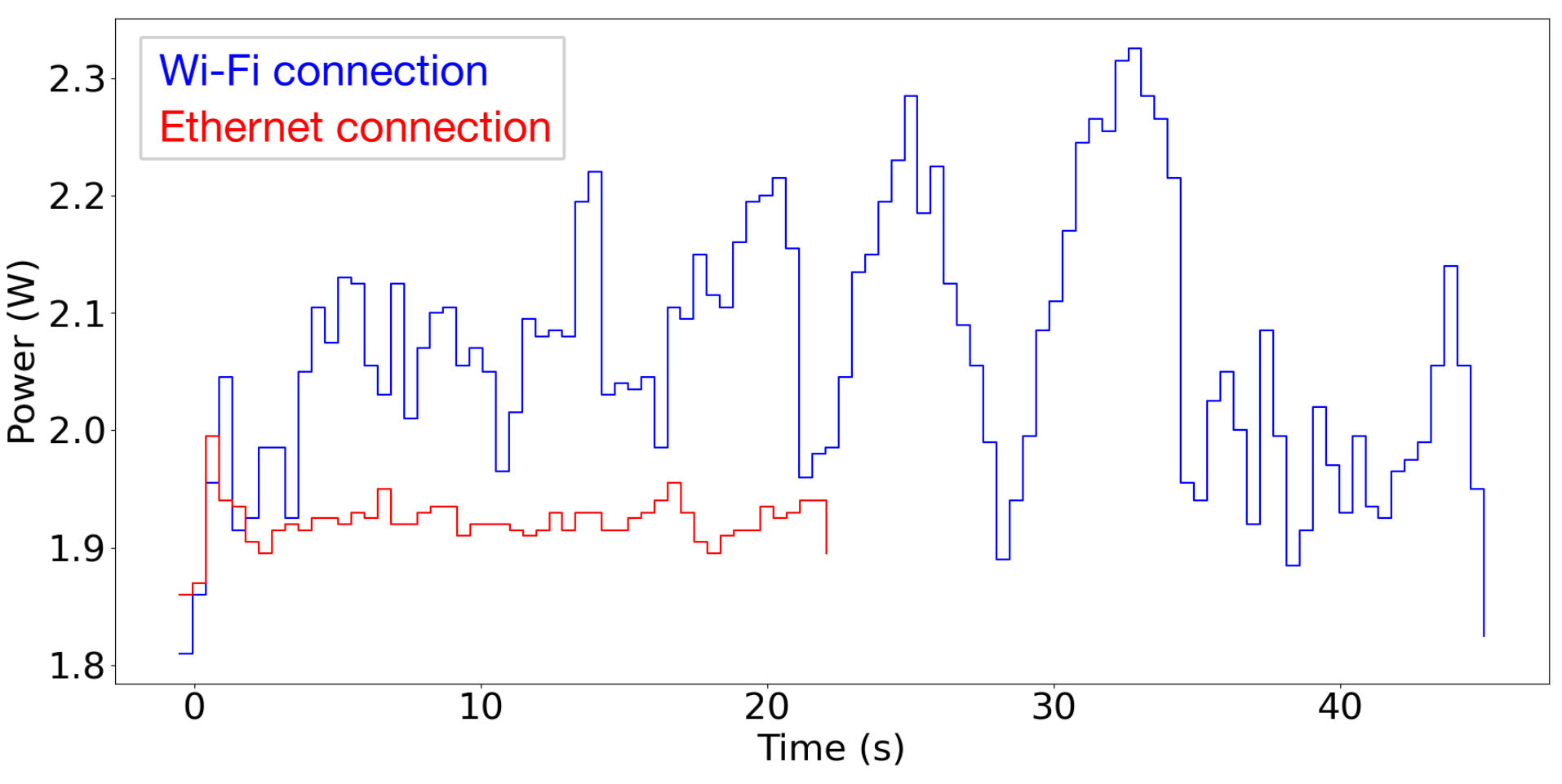}
\caption{Power consumption of the full setup uploading 50MB to a distant server. Two types of connectivity were tested: Wi-Fi (blue) and ethernet (red).}
\label{fig:iperf_upload}
\end{figure}

\paragraph{Results}

The results in Figure \ref{fig:iperf_download} and in Figure \ref{fig:iperf_upload} show the comparison between the types of connection. Although it is obvious that a wired connection is faster than a wireless, downloading data with the first reaches a higher spike of power than the latter, while uploading data show the opposite results.

\begin{table}[htbp] \centering
\caption{Average energy consumed  while performing connectivity tests (transferring 50MB)}
 \resizebox{0.7\columnwidth}{!}{
	\begin{tabular}{ |l|cc|cc|  }
	 \hline
	Network& \multicolumn{2}{c|}{Data-rate} & \multicolumn{2}{c|}{Power consumption}\\
	& \small{download} & \small{upload} & \small{download} & \small{upload} \\
	 \hline
	 Wi-Fi  & 25.7 Mbits/s & 9.5 Mbits/s &  34.0 Joules & 93.9 Joules\\
	 Ethernet & 86.4 Mbits/s & 19.2 Mbits/s &10.8 Joules & 42.5 Joules\\
	 \hline
\end{tabular}
}
\label{tab:iperf}
\end{table}

Table~\ref{tab:iperf} helps to understand what is the limiting factor and the pressure point in each case:
\begin{itemize}
 \item While downloading through Wi-Fi, the speed seems to be limited by the Raspberry Pi's Wi-Fi card.
 \item While downloading through Ethernet, the speed is limited by the Ethernet cable.
 \item While uploading through Wi-Fi, the speed seems to be limited by the Raspberry Pi's Wi-Fi card.
 \item While uploading through Ethernet, the speed is limited by the local network's upload bandwidth. This could explain the rather small increase of power in this case.
\end{itemize}

Table~\ref{tab:iperf} also shows that it is more efficient to use an Ethernet connection while exchanging data. However, while in Idle mode or performing other tasks that do not stress the network, having an Ethernet cable plugged into a Raspberry Pi introduces an extra energy cost of around 0.07W to 0.10W. It represents the energy required to handle the Ethernet peripheral. The longer a Raspberry Pi is performing tasks that do not require the Internet, the better the Wi-Fi connection compared to Ethernet.


%


\section{Beekeeping system in different temperature conditions}
\label{temperature}
Precision beekeeping systems will be deployed on-site in wild zones. The system is supposed to run without interruption outdoors and yearly. How varying temperature conditions can impact the energy consumption of the system?

\begin{figure}[htbp]
\centering
\includegraphics[width=0.7\columnwidth]{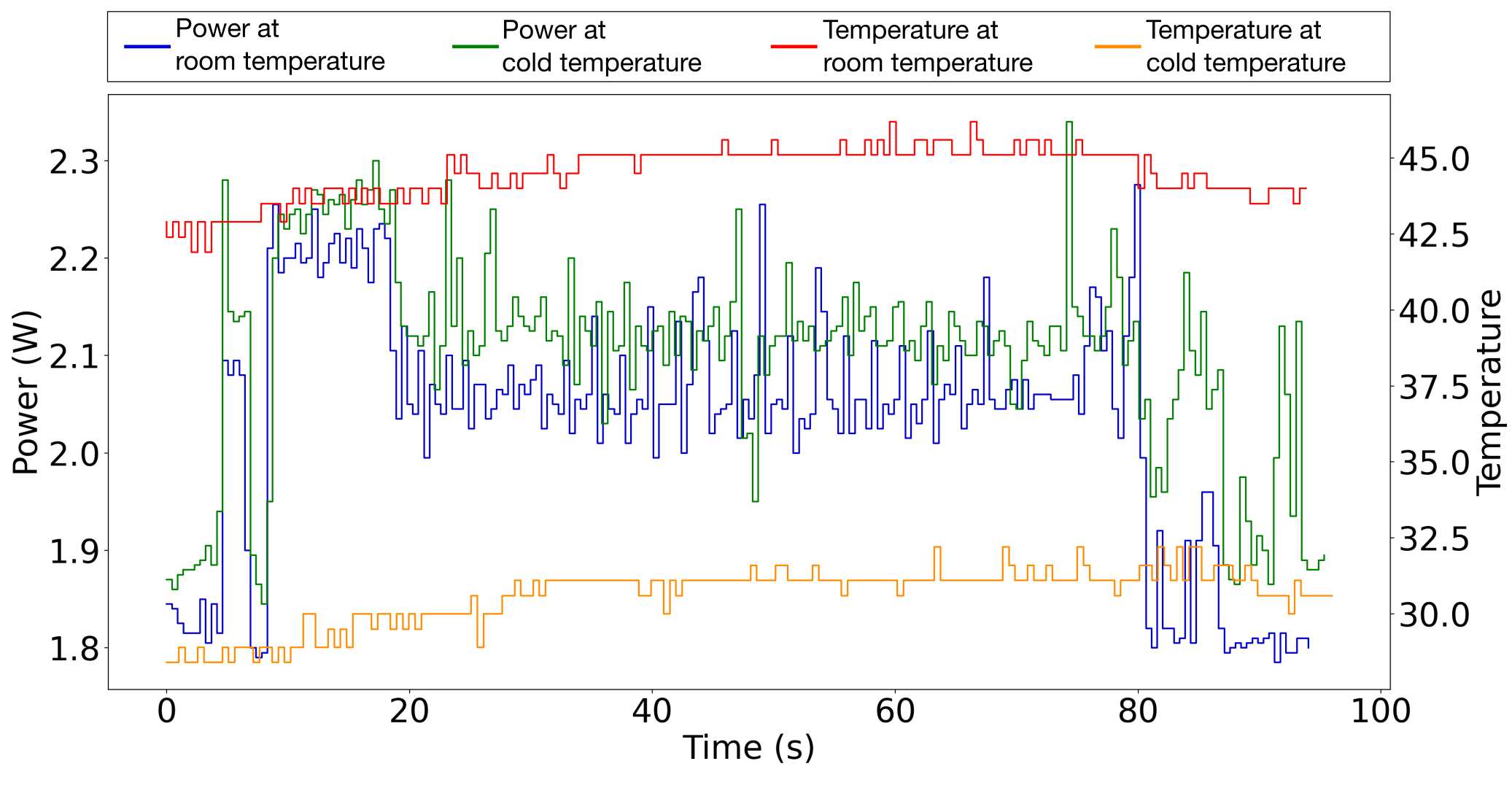}
\caption{Power usage and temperature of the Raspberry Pi chip for one script run  at different temperature conditions}
\label{fig:global_power_temp}
\end{figure}


\paragraph{Experimental Conditions}

The testbed (Figure~\ref{fig:setup}) was placed in two different temperature conditions: at a cold temperature (inside a fridge; between 3\textdegree{}C and 5\textdegree{}C)  and at room temperature (between 19\textdegree{}C and 22\textdegree{}C), all the while maintaining power supply connection.
For each one (Figure~\ref{fig:global_power_temp}), ten iterations of the main script were executed, each of them followed by a 3-minute break, to retrieve a resting state.
Since the pictures captured in an open place (room temperature situation) are different than the ones taken inside a closed dark fridge (cold temperature situation), the size can differ too, leading to different performance for compressing images and converting them into a GIF file. 
To avoid this potential issue, the same batch of images was always chosen to create the GIF file.

\paragraph{Results}
Over 10 iterations of the main script, the average energy consumption is on average 169.6 Joules at room temperature, with a standard deviation of 3.1 Joules (average temperature of Raspberry Pi chip: 45.7\textdegree{}C ; standard deviation: 0.6\textdegree{}C). At cold temperature (between 3\textdegree{}C and 5\textdegree{}C), the average consumption of energy is slightly impacted with an average consumption of 186.4 Joules and a standard deviation of 4.6 Joules (average temperature of Raspberry Pi chip: 30.4\textdegree{}C ; standard deviation: 0.5\textdegree{}C).

Practically, the system (full setup with Raspberry Pi and Arduino plugged in) alternates between a 60-minute OFF mode and the 5-minute data collection window. The latter is a combination between the Raspberry Pi boot up, the execution of the script, and the shutdown. According to the observations in the previous sections, the total cost of energy is 1800 Joules for the 60-minute OFF mode phase and 548 Joules for the 5-minute script window. All in all, that is an average of 2167 Joules per hour, being 120.4 mAh at 5V. In the example of the currently deployed system~\cite{Ammar}, the capacity of the solar-powered battery is 33000 mAh. So if a fully charged battery were to lose its link to its charging source, the system would still function for around 11 days and 10 hours (274 cycles of 1 hour).

\section{Conclusion and Future Work}
\label{conclusion}
This work presents a complete analysis of energy consumption and the thermal performance of a precision beekeeping system.
This is the first step towards optimizing the energy consumption of the deployed services of a connected beehive.


Based on our analysis, it seems like the temperature of the system does not significantly affect the energy consumption of the execution of a script.
The current approach of the Makers' Beehive, which consists of shutting down completely the system and waking it up every hour in order to perform the data collection seems to be an energy efficient solution because the residual consumption of an idle Raspberry Pi 3 has a significant impact on energy consumption.
The consumption of the USB connection of the Arduino sensor node is significant. It adds an extra fixed cost of energy, so future work will explore other ways for the plugging sensor (sensor shield) in order to determine the most energy efficient solution.

Future work will focus on deploying new services of data processing like deep learning applications in the precision beekeeping system. Taking care of the energy consumption is mandatory since there are multiple ways to proceed: either performing the computation on chip (energy-heavy task) and only sending the results (few bytes), or not performing any computation and sending the uncompressed data streams.

\bibliographystyle{siam}
\bibliography{literature}

\end{document}